\documentclass[useAMS,usenatbib,includegraphicx]{mn2e}

\usepackage{graphicx,subfigure}
\newcommand{\msun}{M$_{\odot}$}

\begin{document}

\title[Pulsation in LBVs]{Pulsations as a Driver for LBV Variability}

\author[Lovekin \& Guzik]{C.C.Lovekin$^{1,2}$\thanks{email: clovekin@mta.ca} \& J.A. Guzik$^3$\\
$^1$T-2, Los Alamos National Laboratory, Los Alamos, NM, 87545\\
$^2$Department of Physics, Mount Allison University, Sackville, NB, Canada\\
$^3$Theoretical Design Division, XTD-NTA, MS T086, Los Alamos National Laboratory, Los Alamos, NM, 87545  USA}

\maketitle

\begin{abstract}
Among the most spectacular variable stars are the Luminous Blue Variables (LBVs), which can show three types of variability.  The LBV phase of evolution is poorly understood, and the driving mechanisms for the variability are not known.    The most common type of variability, the S Dor instability, occurs on timescales of tens of years.  During an S Dor outburst, the visual magnitude of the star increases, while the bolometric magnitude stays approximately constant.  In this work, we investigate pulsation as a possible trigger for the S Dor type outbursts.  
We calculate the pulsations of envelope models using a nonlinear hydrodynamics code including a time-dependent convection treatment. We initialize the pulsation in the hydrodynamic model based on linear non-adiabatic calculations.  Pulsation properties for a full grid of models from 20 to 85 \msun were calculated, and in this paper we focus on the few models that show either long-period pulsations or outburst-like behaviour, with photospheric radial velocities reaching 70-80 km/s.  
At the present time, our models cannot follow mass loss, so once the outburst event begins, our simulations are terminated.  Our results show that pulsations alone are not able to drive enough surface expansion to eject the outer layers.  However, the outbursts and long-period pulsations discussed here produce large variations in effective temperature and luminosity, which are expected to produce large variations in the radiatively driven mass-loss rates.  

\end{abstract}


\section{Introduction}
\label{intro}

Luminous Blue Variables (LBVs) are spectacular examples of variable stars,  which can brighten by orders of magnitude on timescales of months to years.  These stars are typically more evolved than the $\beta$ Ceph stars, and undergo periods of outburst, typically characterized by enhanced mass-loss rates.  The most energetic of these eruptions, known as giant eruptions, occur over relatively long timescales, lasting a few tens of years, and are thought to have long periods of dormancy between eruptions \citep{hd94,smithrev}.  The best known examples of these events have occurred in the stars $\eta$ Car and P Cyg.  The total mass ejected in a giant eruption of this kind can be more than 1 \msun.  For example, the great eruption of $\eta$ Car in the mid-nineteenth century may have ejected as much as 10-15 \msun\ \citep{smith03}. During these events, both the temperature and the luminosity of the star increase, typically by 1-3 bolometric magnitudes.  The increase in luminosity can be so large that the stellar luminosity is well above the Eddington limit, which is likely a driving factor in the amount of mass lost.  The amount of brightening can lead to these stars being mistaken for supernovae, and so these stars are sometimes called `supernovae impostors'  \citep[see][]{vandyk05,galyam,smith11,vandyk}.

Although giant eruptions are the most easily visible form of variability in LBVs, S Dor variability is more common.  S Dor variations have a shorter timescale than giant eruptions, lose less mass, and the bolometric luminosity remains unchanged.  The timescale for an S Dor outburst is years to decades.  The quiescent interval between outbursts is highly variable, but the timescale can generally be categorized as either a short timescale, less than 10 years, or a long timescale, greater than 20 years.  These events do not eject mass the way a giant eruption does, but instead the stars undergo an episode of enhanced mass loss, typically ejecting around 10$^{-4}$ \msun\ per event.  During an S Dor episode, the visual magnitude of the star increases, typically by 1-2 magnitudes, while the bolometric magnitude remains the same.  The position of the star moves horizontally in the HR diagram, and instability strips for both the quiescent and outburst phase can be defined observationally.  In this work, we focus on this intermediate type of variability.

LBVs also show a third type of variability, known as microvariability.  These variations are small, typically a few tenths of a magnitude.  They appear to be stochastic, with irregular periods on the order of weeks to months \citep{abolmasov}.  For a more detailed review of LBV properties, see the recent review by \citet{vink12}. 

S Dor variability, as seen in AG Car and S Dor for example, was initially thought to be caused by an increase in mass loss causing the stellar photosphere to be located in a dense outflow (a `pseudo-photosphere').  As a result, the star appeared cooler compared to a `normal' star \citep[e.g][]{davidson}.  Indeed, the mass-loss rates during this type of outburst have been observed to be variable \citep[e.g.][]{stahl01,stahl03}.  
The underlying cause of these variations is not yet understood.  We investigate radial pulsation as a possible origin for the variations, although nonradial pulsations may also play a role.

The evolutionary status of LBVs is still uncertain.  Theoretical work has determined them to be a post-main sequence evolution phase during which massive O stars lose their outer layers before becoming Wolf-Rayet (WR) stars, eventually producing SN \citep[e.g.][]{chiosi}.  Observationally, LBVs are cooler than the core-He burning WR stars.  The most luminous LBVs are brighter than the horizontal part of the Humphreys-Davidson (HD) limit, but hot enough to not violate the limit.  They may be on their first crossing of the HR diagram in a shell H burning or core He burning phase \citep[see, for example][]{hd94}. The envelopes of some LBVs are observed to have enriched He abundances and CNO-cycle processed material, which is thought to indicate that the outer layers of the stars have been removed, revealing material that has previously undergone H burning \citep{najarro97,najarro01}. However, similar enrichment is expected in stars that are on the second crossing of the HR diagram.  It is thought that LBVs occur when massive stars evolving redward encounter the HD limit.  Below this limit, stars are able to continue to the red to become red supergiants.  As the HD limit is an empirical limit, the mass range where stars can become LBVs depends on the stellar models used. Although many explanations have been proposed, there is no accepted physical explanation for the location of the limiting mass or the exact cause of the LBV phenomenon, and the HD limit is determined observationally.

Recent observations have shown that the evolutionary status of LBVs may be more complicated than described above.  Recent downward revisions of the theoretical mass-loss rates indicate that radiatively driven mass loss cannot remove enough material to match the observed masses of WR stars.  If the majority of massive stars are expected to end as WR stars, the most massive stars may thus need a series of giant eruptions during an LBV phase to reach the masses of typical WR stars \citep{hd94, smith06}.  However, the traditional evolution sequence is now being challenged, as there is some evidence that LBVs can undergo giant $\eta$ Car-like eruptions shortly before exploding as either Type IIn or superluminous SN \citep{smith10}.  Classical S Dor variables have not been found to be immediate progenitors of SN.  These problems may be explained if LBVs are actually products of binary evolution, as suggested by \citet{smith14}.  \citet{kotak} have proposed that SNe which show quasi-sinusodial variation in their radio light curves are produced by S Dor type LBV progenitors.  As the mass-loss rate changes during the S Dor cycle, the surrounding circumstellar medium will have a variable density, causing variations in the light curve during the interaction with the SN shock wave.  A better understanding of S Dor variability may help shed light on some of these open questions.

It has been suggested that pulsation could be the origin of S Dor variability.  In particular, the quasi-periodic variability observed in massive stars could be explained by radial strange modes \citep{saio09,dorfi00}.  These modes arise when the pulsating region has much higher radiation pressure than gas pressure, i.e., highly non-adiabatic environments.  Other authors have argued the observed periods are too long to be strange modes, but are compatible with non-radial g modes.  These modes have been shown to be unstable \citep{shibihashi81}, but were initially expected to be too low amplitude to be easily observable.  However, more recent work by \citet{saio} has shown that some low $\ell$ modes with long periods are expected to be observable at the stellar surface. 

In this work, we investigate radial pulsations as a driver of LBV-like variability in stellar models from 20 to 85 \msun.   It has been suggested that $\kappa$-effect pulsations arising from a bump in the opacity caused by Fe ionization near 200\,000K, combined with time-dependent convection could produce a super-Eddington layer, causing the star to become unstable and drive mass loss \citep[see][and references therein]{guzik12,onifer}.  The convective zone in the models we consider can be very dynamic, and grows or shrinks during the pulsation cycle.  The majority of the luminosity is generally still carried by radiation, but the variability in the convective region can produce a super-Eddington region in any of the zones above the 200\,000 K layer.   We have extended previous work by looking at stellar models covering a wide range of masses, including those that are traditionally thought to be too low mass to produce LBVs.

In \S \ref{models} we describe the stellar models, the linear and nonlinear modelling of the pulsations, including time dependent convection, and the methods used to extract periods and growth rates from the nonlinear simulation data.  We have calculated non-linear pulsation periods for a grid of models ranging from 20 to 85 \msun.  Here we focus on those few models which show evidence of outburst-like behaviour or very long quasi-periodic variability.  We discuss the long-period models in \S \ref{longp}, and the behaviour of the models during outburst is discussed in \S \ref{outburst}.  The effects of pulsation on the radiatively driven winds are discussed in \S \ref{winds}, and our results are summarized in \S \ref{conclusions}.

\section{Models}
\label{models}

We use one-dimensional, non-rotating stellar models along an evolutionary sequence from the grid published by \citet{mm94}.  We consider models at metallicities 0.004, 0.008, 0.02 and 0.04, and masses of 20, 40, 60 and 85 \msun.   The models were sampled at intervals along the late main sequence and subgiant branch, typically giving 5-10 models at each mass and metallicity combination.  We used the mass, temperature, luminosity and surface helium abundance of these models to calculate envelope models, consisting of 60 zones.  The mass of each model was adjusted to ensure the bottom of the envelope reached a depth of 1-3 million K.  This temperature ensured our models fully captured the driving region (around 200\,000 K) and the damping region.  Typically, the mass amounted to less than 5 \% of the total mass of the star.  To test the zoning in our models, we calculated a few non-linear models with 120 zones and found that the resulting periods were unchanged to within a few percent.  We also performed a few calculations with 1000 zones in the linear model, and found that the periods typically changed by 1-2 \%.  The normalized work changed by factors of 10-50 \%, although the sign did not change.  

For each envelope model, we calculated the first four radial pulsation frequencies using a linear non-adiabatic pulsation code \citep{cox83}.  The results of this linear pulsation calculation were used as input to a 1D hydrodynamic code \citep[Dynstar,][]{ostlie} with time-dependent convection.  The time-dependent convection implemented in this code is described in more detail below.  Each model was initialized with a single mode at a velocity of 1 km/s.   We have found that the variations discussed here are relatively insensitive to the initial overtone.  For simplicity, we generally show the results for models initialized in the fundamental mode. Our models also neglect rotation, which would also extend the photosphere and decrease the escape velocity.  We hope to address this further in the future.  Once the pulsation has been initialized, the envelope model is evolved hydrodynamically for many pulsation cycles.  

To determine the pulsation period, we fit a simple sine curve to the radial velocity variations.  This method gives us the period and amplitude of the dominant pulsation period.   In practice, many of our light curves appear to be multi-periodic, as shown in Figure \ref{fig:quasiP}, but we consider only the dominant period here.  We also fit the radial velocity variations to estimate a nonlinear growth rate.  This is done by smoothing the light curve using  a moving boxcar filter and fitting the result to an exponential function ($y = ae^{-bx}+c$).  The coefficient of the exponent, $b$, is taken to be the nonlinear growth rate of the pulsation. 

Our long-period and outbursting models show large variations in the photospheric properties, including effective temperature and radius.  Even if the pulsation amplitudes are not large enough for mass to escape from the surface, the large variations in photospheric parameters will lead to large variations in the radiatively driven mass-loss rates. Determination of physical mass-loss rates can be highly uncertain, particularly in pulsating stars such as those discussed here, where the surface properties are changing rapidly over relatively short time scales.  However, we will discuss the variations in the physical parameters that help determine mass-loss rates and what effect the variations in the parameters are likely to have on the resulting mass loss. 

\subsection{Time-Dependent Convection}
\label{tdc}

Standard evolution models with mixing-length theory generally assume a static structure with unvarying thermodynamic quantities in each zone that result in a constant luminosity transport via convection, and complete mixing within this region.  Linear pulsation models have for the most part, adopted either a frozen-in convection approximation, or an instantaneously adapting convection. Reality is likely to be somewhere in-between these approximations.   When a region becomes convectively unstable, there is some finite period of time during which the convective flows become established.  In the nonlinear pulsating models described here, there are regions which periodically become unstable to convection.  In our standard linear pulsation model with frozen-in convection, the convection carries away excess energy according to the hydrostatic equilibrium conditions for the model that do not vary with the phase of the pulsation cycle.  In a model with time-dependent convection, convection cannot immediately carry away the excess energy, which may cause buildup of energy and expansion in these layers. Time-dependent convection, coupled with pulsations can therefore change the structure of envelope models compared to a model in hydrostatic equilibrium assuming either a frozen-in or instantaneously adapting convection approximation.  
 
The model of time-dependent convection used here is based on that outlined by \citet{ostlie}, and is a modification of the \citet{MLT} mixing length theory

In a static model, the luminosity carried by convection is given by 
\begin{equation}
L_{conv} = 4\pi r^2\frac{4T\rho C_p}{g\ell Q}{v_c^o}^3
\end{equation}
where $r$, $T$, p, $\rho$, $C_p$ and $g$ are the local radius, temperature, pressure, density, specific heat and gravity respectively.  The convective velocity, $v_c^o$ is given by
\begin{equation}
v_c^o = \frac{1}{r\sqrt{2}}\frac{g^{1/2}Q^{1/2}\ell}{H_p^{1/2}}f
\end{equation}
where the mixing length, $\ell$ is a variable length parameter, typically of order $H_p$.  We have used a mixing length of 1.5 $H_p$ in these simulations.  The pressure scale height, $H_p$, is defined to be $P/\rho g$, and $Q$ is the density gradient
\begin{equation}
Q = \left(-\frac{\partial log \rho}{\partial log T}\right )_P
\end{equation}
which is exactly 1 for an ideal gas with constant mean molecular weight.   
The parameter $f$ is defined as 
\begin{equation}
f = \left[\sqrt{1  + 4A^2(\nabla-\nabla _{ad})}-1\right]/A
\end{equation}
where 
\begin{equation}
\nabla = \frac{dlogT }{dlogP}
\end{equation}
and 
\begin{equation}
A \equiv \frac{Q^{1/2}C_p\kappa g\rho ^{5/2}\ell ^2}{12\sqrt{2}acP^{1/2}T^3}.
\end{equation}
$\nabla-\nabla_{ad}$ is the superadiabatic gradient, and $\kappa$ is the local opacity of the material.  

In the time-dependent model implemented in this work, the convective velocity is interpolated using the previous two time steps and the instantaneous value calculated using standard mixing length theory.  The time-dependence of the convective velocity in zone $i$ at timestep $n$ is given by 
\begin{eqnarray}
v^n_{c,i} & =& \frac{(t'-t_{n-1})(t'-t_{n-2})}{(t_n-t_{n-1})(t_n-t_{n-2})}v^o_{c,i}\\
& & + \frac{(t'-t_n)(t'-t_{n-2})}{(t_{n-1}-t_n)(t_{n-1}-t_{n-2})}\overline{v}^{n-1}_{c,i}\\
& & + \frac{(t'-t_n)(t'-t_{n-1})}{(t_{n-2}-t_n)(t_{n-2}-t_{n-1})}\overline{v}^{n-2}_{c,i}
\end{eqnarray}
where 
\begin{equation}
t' \equiv t_{n-1}+\tau(t_n-t_{n-1})
\end{equation}
and
\begin{equation}
\tau \equiv \frac{v^n_{c,i}(t-t_{n-1})}{\ell}l_{fac}
\end{equation}
where $\tau$ is between 0 and 1, and $l_{fac}$ is a free parameter that determines the fraction of a mixing length that the convective eddy can travel in one time step.  In our models, $l_{fac} = 1.0$.  Hence, there is a time lag included in the equations, and there can be a delay before the convective flow begins or ends in unstable regions.  

This treatment of convection also includes non-local effects, included using a spatial average over neighbouring zones, as these are assumed to have some influence on local conditions.  Once all these effects are taken into account, the convective velocity is given by
\begin{equation}
\label{vcon}
\overline{v_{c,i}^n} = a_{i-1}v_{c,i-1}^{n-1} + a_{i+1}v_{c,i+1}^{n-1} + (1-a_{i-1} - a_{i+1})v_{c,i}^n
\end{equation}
where $v_{c,i}^n$ is the convective velocity from the time averaging for the current time step ($n$) and zone ($i$).  The weighting factors in Equation \ref{vcon}, $a_k$ for zone $k$, are given by
\begin{equation}
a_k = a_{fac}\left(1-\frac{|r_k-r_i|}{\ell}\right)
\end{equation}
where $k$ is either $i-1$ or $i+1$, $a_k$ is constrained to be between 0 and an averaging factor, $a_{fac}$, and $a_{fac}$ is a free parameter of order unity.  In our models, $a_{fac}$ = 0.8.

\section{Long-Period Models}
\label{longp}

\begin{figure}
\center
\includegraphics[width=0.5\textwidth]{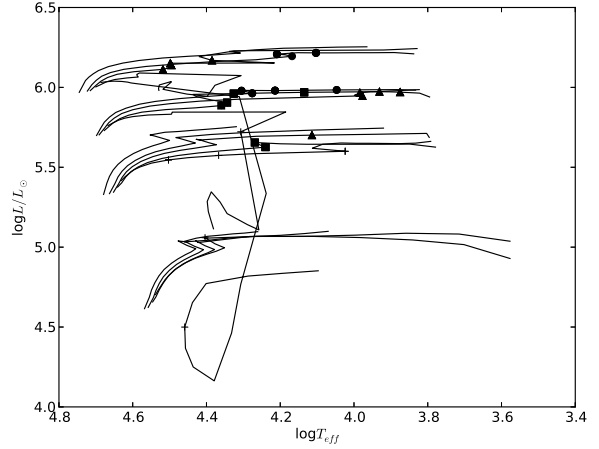}
\caption{\label{fig:long}Evolution tracks from \citet{mm94} for 20, 40, 60, and 85 \msun\ models with metallicities Z=0.004 (circles), 0.008 (triangles), 0.02 (squares) and 0.04 (+).  Points indicate the models which were fit with periods of 20 d or longer.  }
\end{figure}

\begin{figure}
\center
\includegraphics[width=0.5\textwidth]{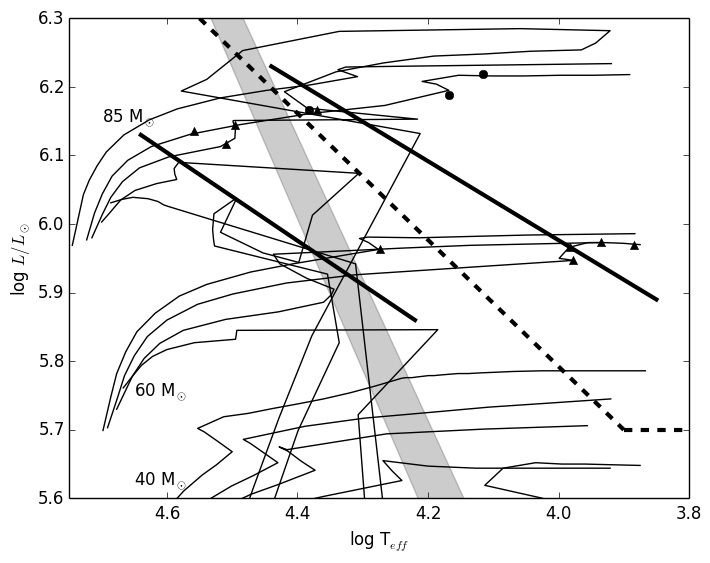}
\caption{\label{fig:longmods}The location of 60 and 85 \msun\ models that show long periods (triangles), defined here as being longer than $\sim$ 50 days, or outbursts (circles).  At each mass, metallicity increases from top to bottom.  Parts of the 40 \msun\ tracks are also shown on this plot, but as none of the 40 \msun\ models show long periods, they are not discussed here.  Many of these long-period models show variation that is quasi-periodic, as well as having multiple periods.  The thick solid lines show the location of the instability strip for the long-period models, the thick dashed line shows the approximate location of the Humphreys-Davidson limit, and the grey area shows the approximate location of the S Doradus instability strip.  Our instability strip is clearly much larger, and extends to cooler temperatures than the classical S Dor instability strip.}
\end{figure}

\begin{figure}
\center
\includegraphics[width=0.5\textwidth]{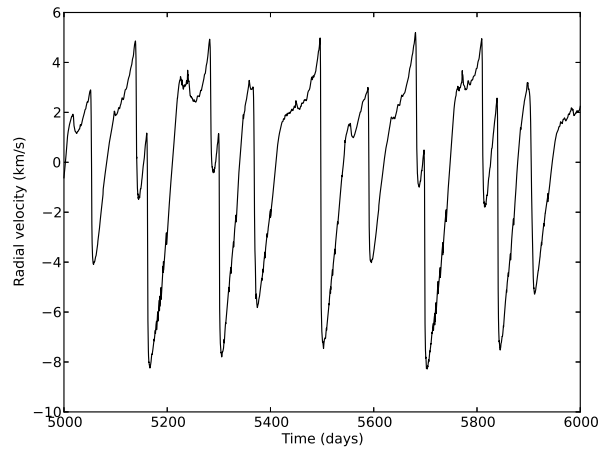}
\caption{\label{fig:quasiP}The photospheric radial-velocity curve for a 60 \msun\ model at Z = 0.008.  Negative radial velocities correspond to an outwardly moving photosphere.  This model shows long-period variability over thousands of days, but is clearly multi periodic, and the dominant features are only quasi-periodic.  }
\end{figure}

\begin{figure}
\center
\includegraphics[width=0.5\textwidth]{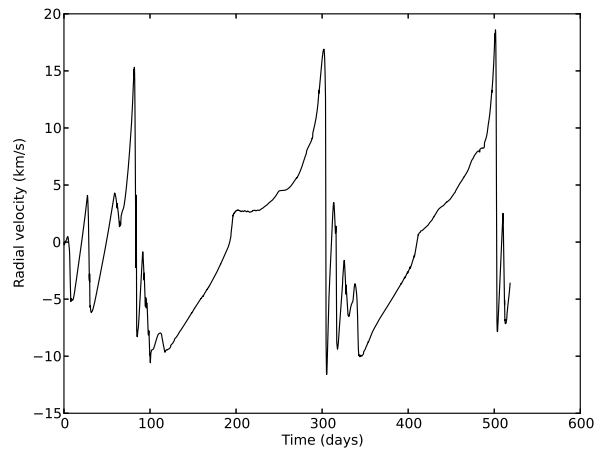}
\caption{\label{fig:longrv}The photospheric radial-velocity curve for an 85 \msun\ model with Z = 0.004.  Negative radial velocities correspond to an outwardly moving photosphere.  Unlike the outbursting model described below, this model is still on the main sequence.  Although the radial velocities remain relatively small, the long timescales involved mean that the radius and temperature still change significantly.  The effects on the mass-loss rates for this model are discussed below.}
\end{figure}

We have examined a subset of models that were found to have long periods based on the fits to the radial-velocity curves.  Long-period models in this paper are arbitrarily defined as those that have fitted periods of 20 days or longer.  Models which are fit by very long periods tend to fall into three basic categories:  those that are truly long period (e.g., Figures \ref{fig:quasiP} and \ref{fig:longrv}),  those that undergo outburst-like events (e.g., Figure \ref{fig:rv}), and models with light curves that are too short to show periodicity. This third class consists of models that failed to run for extended periods of time for numerical reasons. Models with periods longer than 20 d are indicated in Figure \ref{fig:long}, with circles, triangles, squares and plus signs indicating metallicities of 0.004, 0.008, 0.02 and 0.04 respectively.  We have examined these models in more detail to determine which models show long-period pulsation.  The results are shown in Figure \ref{fig:longmods}, which shows the location of models with long periods (triangles) and outbursts (circles).  The resulting instability strip is a small fraction of the unstable models, encompassing the post-main sequence of the 60 and 85 \msun\ evolution tracks.  Interestingly, we find pulsating models to the red of the Humphreys-Davidson limit (dashed line on Figure \ref{fig:longmods}), although observationally, stars are not expected to be stable in this region.  It is also interesting to note that given the distribution of outbursting and long-period models, the mechansim does not seem to be related to the bi-stability limit ($\sim$ 20\,000 K), as has been suggested to be the origin of the LBV eruptions \citep[][and references therein]{vink01}.  Although pulsations in this region of the HR diagram are driven by an iron ionization zone around 200\,000 K, and hence depend somewhat on the photospheric opacities, the long period and outburst-like behaviour seen here are driven by the interaction with the time-dependent convection, which is relatively independent of the opacities.  We compared the periods calculated using both a linear non-adiabatic code and a non-linear adiabatic hydrodynamic code, and found that the periods in the models considered here could be hundreds of times longer than predicted by the linear code when time-dependent convection is included.  Without the effects of time-dependent convection, the pulsation periods in these models would be much shorter and the types of behaviour discussed here would not be present. The long-period instability in our models is restricted to the 60 and 85 \msun\ models found on the post-main sequence.  Long-period modes are not found in the most metal-rich models.  As shown in Figure \ref{fig:longmods}, the metal-rich models lose enough mass that the model does not evolve towards the red, and as a result avoids the instability region near the HD limit.  

Many of the long-period models show very complex multi-periodic or quasi-periodic behaviour.  A typical example is shown in Figure \ref{fig:quasiP} for a 60 \msun\ model with Z = 0.008.  This model is relatively cool for a star of this mass, with log $T_{\rm eff}$ = 3.977 and log $L/L_{\odot}$ = 5.947.  
The dominant variability in this light curve ranges from 170 - 280 d, and the pulsation pattern remains stable over nearly 15000 days.  The photospheric radius of this star varies from about 350 to 650 R$_{\odot}$ over the pulsation period, corresponding to a temperature variation of about 2500 K, and a magnitude variation of  0.3.  

As another example, the light curve for an 85 \msun, Z = 0.004 model still on the main sequence is shown in Figure \ref{fig:longrv}.  This model shows periodic behaviour, with a period of approximately 200 d.  The pulsation in this case has a much larger amplitude.  The change in radius over the pulsation cycle is 200 R$_{\odot}$ to 800 R$_{\odot}$, corresponding to a temperature difference of 8000 K.  

\begin{figure}
\center
\includegraphics[width=0.5\textwidth]{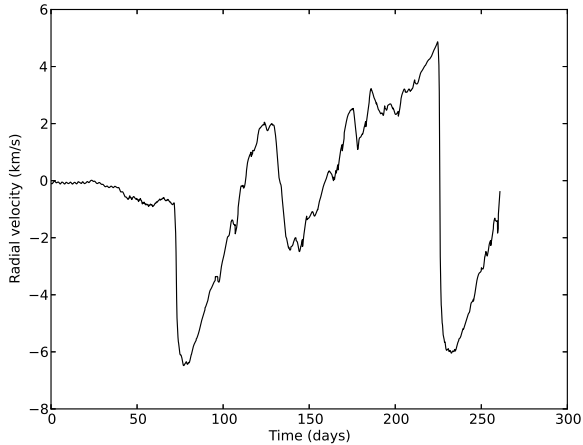}
\caption{\label{fig:aperiodic}The photospheric radial-velocity curve for a 60 \msun\ model at Z = 0.008 showing small amplitude aperiodic variations. Negative radial velocities correspond to an outwardly moving photosphere.  }
\end{figure}

Although some of our models show stable, semi-regular pulsation as shown in Figure \ref{fig:quasiP} and Figure \ref{fig:longrv}, many more show variability that is much more irregular.  The most evolved models in our sample often show aperiodic variation, as shown in Figure \ref{fig:aperiodic} for a 60 \msun\ model at Z = 0.008.  For these models, although there is clearly a significant variability, there is no obvious periodicity on the timescales shown here.  

\section{Outbursting Models}
\label{outburst}

We have classified the outbursts shown in our models into two categories.  The first type are major outbursts, which have peak surface expansion velocities of 50-80 km\ s$^{-1}$ and are indicated by circles on Figure \ref{fig:longmods}.  The second category is the minor outbursts, which show the same general behaviour as the major outbursts, but with expansion velocities limited to 20-30 km s$^{-1}$, which in many cases is not much larger than the underlying pulsation amplitudes.  The minor events have light curves that are similar in shape to the major outburusts and appear to be more common than the major outbursts, which appear in only a few models.  To investigate the mechanism for both major and minor outbursts, we follow the properties in five zones in the interior of our models.  These zones all show similar behaviour, from the hot regions at the bottom of the model to quite close to the surface.  

An analysis of the outbursts indicates that the major and minor outbursts, although superficially similar, are driven by two different mechanisms.  The minor outbursts tend to occur slightly later in the simulation than the major outbursts.  The phase of most rapid outward expansion occurs 10-15 days after the simulation starts, and is usually part of an outward expansion trend.  Shortly before the peak expansion, there is a drop in the radiative luminosity carried through each zone, with a corresponding decrease in the Eddington luminosity (L$_{Edd}$), as calculated using the opacity in each zone.  As the surface continues to expand, the convective motions begin and the convective luminosity increases in the zones we tracked.  As the star stabilizes and the expansion slows, the convection ceases to carry a significant portion of the stellar luminosity and the convective velocities decrease.

Major outbursts, on the other hand, occur earlier in the simulation, typically around 5 days.  In this case, the interior zones are initially convective, and the convective velocities are large.  Early on, the convective velocities begin to decrease rapidly.  This causes a rapid decrease in the convective luminosity and a corresponding spike in the radiative luminosity, causing it to increase well above the Eddington limit at the surface.  This spike is first seen in the deepest zone we tracked, but quickly moves out to the surface.  The super-Eddington radiative luminosity drives the surface to expand rapidly until the convective velocity and luminosity increase as the star stabilizes.  The peak expansion velocities in these models are between 50 and 80 km\ s$^{-1}$, which although rapid, are still significantly less than the escape velocity for these models.  As such, these outbursts are unable to explain major LBV eruptions such as $\eta$ Car.  However, as discussed below, the major outbursts in our models are likely to drive significant mass loss through an increase in the radiataively driven mass-loss rates.  

\begin{figure}
\center
\includegraphics[width=0.5\textwidth]{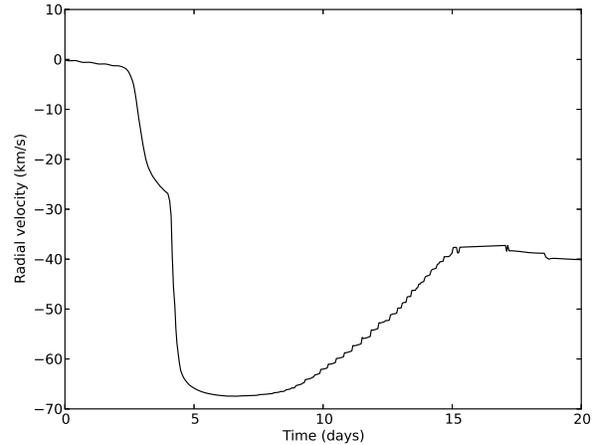}
\caption{\label{fig:rv}The first few days of the photospheric radial-velocity curve for an 85 \msun\ model at Z = 0.004.  Negative radial velocities correspond to an outwardly moving photosphere.  Soon after the simulation is initialized, the surface expands rapidly.  Although the speeds are not high enough to reach the escape velocity and eject the surface layers, it is possible that the wind mass-loss rates (discussed below) will increase significantly.}
\end{figure}

Which models show outbursts is both mass and metallicity dependent.  At high mass, the more metal-poor models are more likely to show an outburst.  As the metallicity increases, outbursts are seen in lower-mass models, with one even appearing in a 20 \msun\ model at Z = 0.04.  An example of an outburst-like event is shown in Figure \ref{fig:rv} for an 85 \msun\ model with Z = 0.004.  This model is actually located just after core hydrogen exhaustion, as the star begins to evolve rapidly toward the red, and has an initial radius of about 230 $R_{\odot}$.  During the outburst, the radius increases to a maximum of 1160 $R_{\odot}$, which corresponds to significant changes in surface temperature. Just before the star begins to expand, the photospheric temperature increases by nearly 15\,000 K, to reach a temperature of 27\,000 K before rapidly dropping to below 10\,000 K during the peak of the outward expansion. This model is initially convective, but the envelope convection shuts off at around 2 days, driving this increase in temperature.  Currently, our Lagrangian hydrodynamics code does not allow mass to be lost from the surface of the model, so once the mass-loss rate becomes significant, the properties of the outer zones of our simulation are not calculated accurately.  At present, we have no way of determining whether these events could be periodic, or what the effect on the subsequent evolution of the model could be.
  
\section{Wind Mass-Loss Rates}
\label{winds}

During the outburst-like events and high-amplitude, long-period models described above, the stellar temperatures and luminosities vary over relatively short timescales. As a result, these stars might be expected to have highly variable mass-loss rates.  The current version of our hydrodynamic code does not allow us to include the effects of mass loss, either from high wind mass loss or from the ejection of material during an outburst.  A revised version of the hydrodynamic code used here is in progress, so we plan to include the effects of mass loss in the future.  Common prescriptions for calculating mass-loss rates, such as that of \citet{vink01}, are not expected to be applicable in these stars, where the variations are on a very short timescale, and the luminosities are high enough that mass loss is largely driven by continuum radiation.  Although we cannot calculate the exact mass-loss rates for these models, we can examine the variations in temperature and luminosity (expressed here as bolometric magnitude) to make qualitative estimates of the mass loss.  

\begin{figure}
\center
\includegraphics[width=0.5\textwidth]{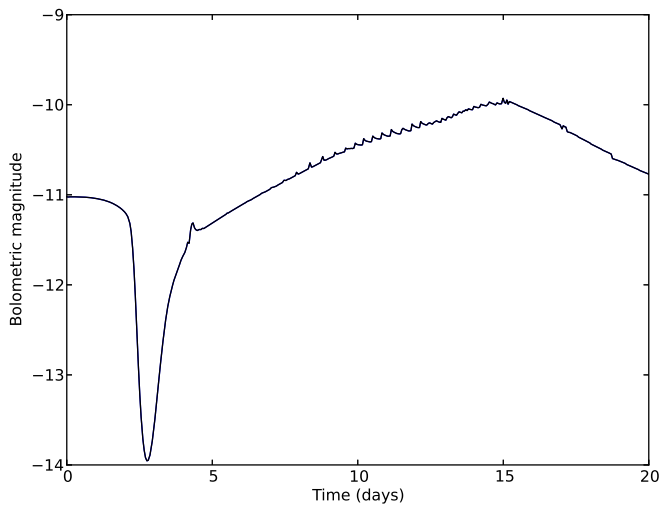}
\caption{\label{fig:outburst_l}The first 20 days of the bolometric lightcurve for the hydrodynamical model of an 85 \msun\ post-main sequence star with Z = 0.004. The sharp increase in brightness is driven by a brief super-Eddington phase at around 2 days.  }
\end{figure}

\begin{figure}
\center
\includegraphics[width=0.5\textwidth]{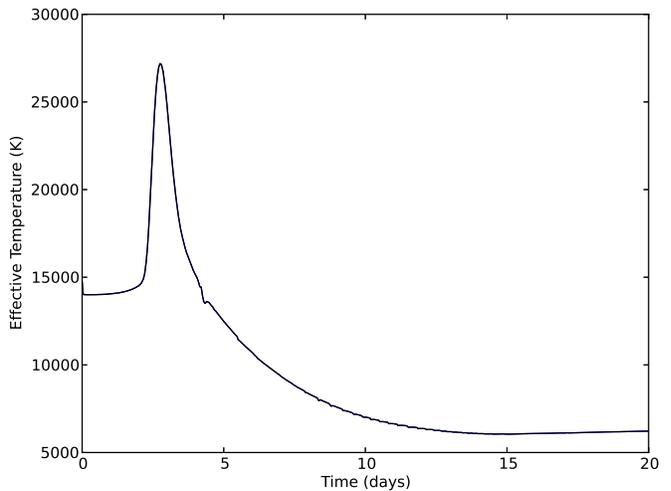}
\caption{\label{fig:outburst_t}The first 20 days of the effective temperature of the hydrodynamical model of an 85 \msun\ post-main sequence star with Z = 0.004.  The increase in temperature corresponds to the increase in brightness seen in Figure \ref{fig:outburst_l}, and is followed by rapid cooling as the star expands. }
\end{figure}

First, we consider a model that shows outburst-like behaviour, namely an 85 \msun\ model on the post-main sequence, with Z = 0.004.  The first few days of the radial-velocity data are shown in Figure \ref{fig:rv}.  The corresponding luminosity and effective temperature variations are shown in Figure \ref{fig:outburst_l} and Figure \ref{fig:outburst_t} respectively. The luminosity from the surface zone of this model is initially slightly above the Eddington limit ($L/L_{Edd} \sim 1.4$), but briefly increases to 21 times the Eddington limit, which causes the star to brighten by 3 magnitudes and drives significant expansion of the star. During this brightening phase, the temperature also increases significantly, from 13\,800 K to 27\,000 K, before cooling off as the star expands.  The luminosity remains above the Eddington limit until approximately 5 days after the start of the simulation, by which time the effective temperature has decreased to approximately 11\,000 K, and is still falling.  In deeper layers of the star, the luminosity becomes super-Eddington slightly later and does not reach as high luminosities, peaking at $\sim 13.5$ times the Eddington luminosity, but remains above the Eddington limit until approximately 7.5 days after the start of the simulation.  It is this super-Eddington layer that appears to drive the expansion of the star seen in Figure \ref{fig:rv}.  During the brief time this model is extremely super-Eddington, the continuum-driven mass-loss rates can be expected to be very high.  High mass-loss rates are also expected to be associated with the very bright, high-temperature phase directly before expansion begins.     

\begin{figure}
\center
\includegraphics[width=0.5\textwidth]{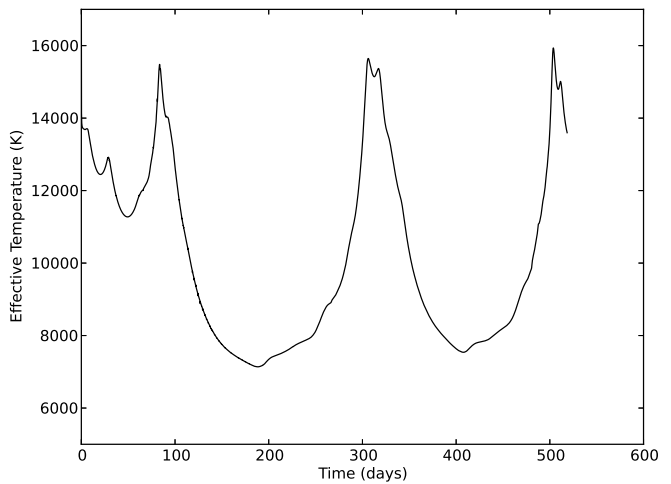}
\caption{\label{fig:longp_t}The variation in effective temperature calculated for an 85 \msun\ main-sequence model with Z = 0.004.  The radial-velocity curve for this model is shown in Figure \ref{fig:longrv}.  The pulsation in this case drives changes in the effective temperature of $\sim$ 8000 K over the approximately 200 d pulsation cycle. }
\end{figure}

\begin{figure}
\center
\includegraphics[width=0.5\textwidth]{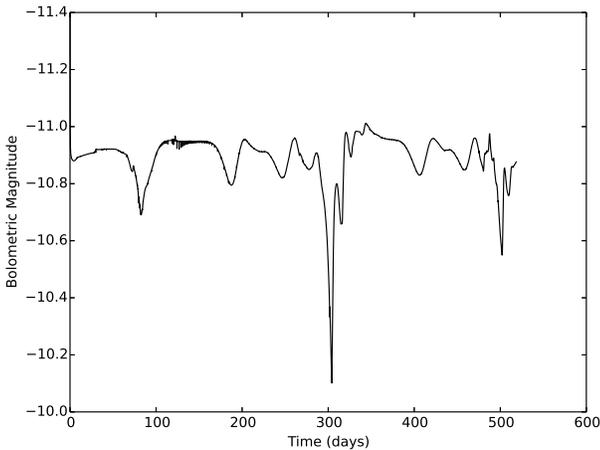}
\caption{\label{fig:longp_m}The bolometric light curve calculated for an 85 \msun\ main-sequence model with Z = 0.004.  The radial-velocity curve for this model is shown in Figure \ref{fig:longrv}.  Estimates of the mass-loss rates show changes of nearly four orders of magnitude over the course of the pulsation cycle. The sharp increase in magnitude at 300 days corresponds to the minimum radius of the pulsation cycle, when the stellar radius is only $\sim$ 125 R$_{\odot}$, down from a maximum of more than 800 R$_{\odot}$.  The initial radius of this model is approximately 220 $R_{\odot}$.}
\end{figure}

The temperature variation and bolometric light curve for the 85 \msun\ long-period model discussed above (Figure \ref{fig:longrv}) are shown in Figure \ref{fig:longp_t} and Figure \ref{fig:longp_m} respectively.  As described above, the long timescales of the variation produce large changes in the radius (600 R$_{\odot}$) and temperature (8000 K).  As a result of these changes, the mass-loss rate also varies dramatically over the pulsation cycle.  Simple calculations using the surface parameters and the \citet{vink01} mass-loss prescription show variations in the mass-loss rate of nearly four orders of magnitude.  Although the mass loss in these models is certainly more complicated than can be described by the \citet{vink01} rates, this estimate does illustrate the variation in mass loss that can be expected in stars undergoing large scale pulsations.  

\section{Conclusion}
\label{conclusions}

In this work, we have calculated pulsations in massive stellar models in order to gain a better understanding of S Dor type LBVs.  Our models cover a wide range of masses, and include metallicities typical of both the Galaxy and the Magellanic Clouds.  Our models include time-dependent convection, which has been proposed to produce super-Eddington regions within the star, driving mass loss \cite[see][and references therein]{guzik12}.  For each model, we calculate a linear, non-adiabatic frequency which is used to initialize a hydrodynamic calculation.  Periods and growth rates are found by fitting the radial-velocity curves produced by the hydrodynamic calculations.  Using both sets of results, we have determined an approximate instability strip, which is found to be slightly hotter and less luminous than the observed location of LBVs.

A few of our models are sufficiently unstable to undergo outburst-like events.  The limitations of our method prevent us from following the model through outburst and recovery, as we do not include mass loss in our models.  Although these outbursts are not large enough to eject matter, the change in surface properties are expected to produce very large changes in mass-loss rates, perhaps comparable to mass-loss rates seen in S Dor outbursts.  Massive models with long, high amplitude pulsations also show large variations in surface temperature and magnitude, which likely translates to large variations in the mass-loss rate.  The significant changes in mass-loss rate can be expected to translate into a variable density circumstellar medium, and may be able to explain SNe with quasi-sinusoidal radio light curves.

\section*{Acknowlegements}
The authors would like to thank Roberta Humphreys for her many helpful comments, which have greatly improved this manuscript.  The authors would also like to thank Bob Deupree for his feedback on an early version of this paper.  This work was performed for the U. S. Department of Energy by Los  Alamos National Laboratory under Contract No. DE-AC52-06NA2-5396.


\begin{thebibliography}{}

\bibitem[Abolmasov(2011)]{abolmasov}Abolmasov, P.  2011.  New Astronomy, 16, 421
\bibitem[Blomme et al.(2011)]{blomme}Blomme, R., Briquet, M., Degroote, P., et al.\ 2011.  in {\it Four decades of research on massive stars}, PASP, eds C. Robert, N., St-Louis \& L. Drissen.  
\bibitem[B\"ohm-Vitense(1958)]{MLT}B\"ohm-Vitense, E.  1958.  Zs.F.Ap., 46, 108
\bibitem[Chiosi \& Maeder(1986)]{chiosi}Chiosi, C. \& Maeder, A.  1986.  ARA\&A, 24, 329
\bibitem[Cox(1980)]{cox}Cox, J.P.  1980.  {\it Theory of Stella Pulsation}.  Princeton University Press, p. 10
\bibitem[Cox \& Ostlie(1993)]{cox93}Cox, A.N. \& Ostlie, D.A.  1993.  Ap\&SS, 201, 311
\bibitem[Cox et al.(1983)]{cox83} Cox, A.~N., Vauclair, S., 
\& Zahn, J.~P.\ 1983, Saas-Fee Advanced Course 13: Astrophysical Processes in Upper Main Sequence Stars
\bibitem[Davidson(1987)]{davidson}Davidson, K.  1987.  ApJ, 317, 760
\bibitem[Dorfi \& Gautschy(2000)]{dorfi00}Dorfi, E.A. \& Gautschy, A.  2000.  ApJ, 545, 982
\bibitem[Gal-Yam et al.(2007)]{galyam}Gal-Yam, A., Leonard, D.C., Fox, D.B., et al.  2007.  ApJ, 656, 372
\bibitem[Gautschy \& Glatzel(1990)]{gg}Gautschy, A. \& Glatzel, W.  1990.  MNRAS, 245, 597
\bibitem[Guzik \& Lovekin(2012)]{guzik12}Guzik, J.A. \& Lovekin, C.C.  2012.  AstRev, 7, 13, arXiv:1402.0257
\bibitem[Humphreys \& Davidson(1994)]{hd94}Humphreys, R.M. \& Davidson, K.  1994.  PASP, 106, 1025
\bibitem[Kotak \& Vink(2006)]{kotak}Kotak, R. \& Vink, J.S.  2006.  A\&A, 406, L5
\bibitem[McNamara et al.(2012)]{mcnamara}McNamara, B.J., Jackiewicz, J., \& McKeever, J.  2012.  AJ 143, 101
\bibitem[Meynet et al.(1994)]{mm94}Meynet, G., Maeder, A., Schaller, G., Schaerer, D. \& Charbonnel, C.  1994.  A\&AS, 103, 97
\bibitem[Moffat(2010)]{moffat}Moffat, A.F.J.  2010.  HiA, 15, 366
\bibitem[Najarro(2001)]{najarro01}Najarro, F.  2001. P Cygni 2000: 400 Years of Progress, 233, 133 
\bibitem[Najarro et al.(1997)]{najarro97}Najarro, F., Hillier, D.J. \& Sthal, O.  1997.  A\&A, 326, 1117
\bibitem[Onifer \& Guzik(2008)]{onifer}Onifer, A.J. \& Guzik, J.A.  2008.  in {\it Massive Stars as Cosmic Engines}, edited by F. Bresolin, P.A. Crowther \& J. Puls, Proceedings of IAU Symposium 250, p 83.  
\bibitem[Ostlie(1990)]{ostlie}Ostlie, D.A.  1990.  in {\it Numerical Modeling of Nonlinear Stellar Pulsations Problems and Prospects}, ed. J.R. Buchler, Kluwer Academic, Dordrecht, p89
\bibitem[Pamyatnykh(1999)]{pam99}Pamyatnykh, A.A.  1999.  AcA, 49, 119
\bibitem[Saio(2011)]{saio}Saio, H.  2011.  MNRAS, 412, 1814
\bibitem[Saio(2009)]{saio09}Saio, H.  2009.  CoAst, 158, 245
\bibitem[Shibahashi \& Osaki(1981)]{shibihashi81}Shibahashi, H. \& Osaki, Y.  1981.  PASJ, 33, 427
\bibitem[Smith(2010)]{smith10}Smith, N., in {\it Hot and Cool:  Bridging gaps in massive star evolution}, eds. C. Leithere, P. Bennett, P. Morris, J. van Loon.  ASP Conference Series 425, 63
\bibitem[Smith(2011)]{smithrev}Smith, N.  2011.  in {\it Active OB stars: structure, evolution, mass loss, and critical limits}.  Proceedings of the IAU Symposium 272, 571
\bibitem[Smith et al.(2003)]{smith03}Smith, N., Gehrz, R.D., Hinz, P.M, Hoffmann, W.F., Hora, J.L., Mamajek, E.E. \& Meyer, M.R.  2003.  ApJ, 125, 1458
\bibitem[Smith \& Owocki(2006)]{smith06}Smith, N. \& Owocki, S.P.  2006.  ApJ, 645, L45
\bibitem[Smith \& Tombleson(2014)]{smith14} Smith, N., \& Tombleson, R.\ 2014, arXiv:1406.7431 
\bibitem[Smith et al.(2011)]{smith11}Smith, N., Li, W.  Silverman, J.M., Ganeshalingam, M. \& Filippenko, A.V.  2011.  MNRAS, 415, 773
\bibitem[Stahl et al.(2001)]{stahl01}Stahl, O., Jankovics, I., Kovacs, J., Wolv, B., Schmutz, W., Kaufer, A., Rivinius, T.,  \& Szeifert, T.  2001.  A\&A, 375, 54
\bibitem[Stahl et al.(2003)]{stahl03}Stahl, O., et al.  2003.  A\&A, 400, 279
\bibitem[van Dyk(2005)]{vandyk05} van Dyk, S.~D.\ 2005, in {\it The Fate of the Most Massive Stars}, eds R. Humphreys \& K. Stanek.  ASP Conference Series 332, 47 
\bibitem[van Dyk \& Matheson(2012)]{vandyk}van Dyk, S.D. \& Matheson, T.  2012.  ``The Supernova Impostors," in {\it Eta Carinae and the Supernova Impostors}, Springer Astrophysics and Space Science Library, 384, 275
\bibitem[van Genderen et al.(2004)]{vangenderen04}van Genderen, A.M., Sterken, C. \& Jones, A.F.  2004.  A\&A, 419, 667
\bibitem[Vink(2012)]{vink12}Vink, J.S.  ``Eta Carinae and the Luminous Blue Variables," in {\it Eta Carinae and the Supernova Impostors}, Springer Astrophysics and Space Science Library, 384, 221
\bibitem[Vink et al.(2001)]{vink01}Vink, J.S., de Koter, A. \& Lamers, H.J.G.L.M.  2001.  A\&A, 369, 574
\end{thebibliography}
\end{document}